# Antiferromagnetic Chiral Bobber Formation and Topological Proximity Effect in MnBi$_2$Te$_4$

Y. Xu[1], D. Kurebayashi[2], D. Zhang[1], P. Schoenherr[1], L. Li[3], Z. Yue[4]*, W. J. Ren[5], M.-G. Han[6], Y. Zhu[6], Z. Cheng[7], X. Wang[7], Oleg A. Tretiakov[2]*, and J. Seidel[1]*

[1] *School of Materials Science and Engineering, UNSW Sydney, Sydney 2052, Australia*

[2] *School of Physics, UNSW Sydney, Sydney 2052, Australia*

[3] *Key Laboratory of Quantum Materials and Devices of Ministry of Education, School of Physics, Southeast University, Nanjing 211189, China*

[4] *School of Artificial Intelligence Science and Technology, University of Shanghai for Science and Technology, Shanghai 200093, China; Institute of Photonic Chips, University of Shanghai for Science and Technology, Shanghai 200093, China*

[5] *Shenyang National Laboratory for Materials Science, Institute of Metal Research, Chinese Academy of Sciences, Shenyang, 110016, China*

[6] *Condensed Matter Physics and Materials Sciences Department, Brookhaven National Laboratory, Upton, NY 11973, USA*

[7] *Institute for Superconducting and Electronic Materials, Institute for Innovative Materials, University of Wollongong, Wollongong, NSW 2522, Australia*

*Corresponding authors: zengjiyue@usst.edu.cn, o.tretiakov@unsw.edu.au, jan.seidel@unsw.edu.au

**Abstract**

With topological materials being billed as the key to a new generation of nanoelectronics via either functional real-space topological structures (domain walls, skyrmions etc.) or via momentum-space topology (topological insulators), tailored and controllable topological properties are of paramount significance, since they lead to topologically protected states with negligible dissipation, enabling stable and non-volatile information processing. Here, we report on the evolution of topological magnetic textures in the proximity of other topological defects, i.e., antiferromagnetic domain walls in the topological insulator MnBi$_2$Te$_4$. The transition from the antiferromagnetic ground state to a canted antiferromagnetic state at finite magnetic fields is accompanied by the formation of chiral bobbers – bulk-terminated topological defects adjacent to the domain walls in this system, leading to a topological proximity effect.

**Introduction**

Magnetism in topological insulators breaks time-reversal symmetry, introduces gaps at the Dirac points on topological surface states, and enhances Berry curvature occurrence (nonzero Chern numbers), which allows for novel real-space topological quantum states,[1-8] the quantum anomalous Hall effect (QAHE),[2-4] axion insulator states,[2-6] and the formation of topologically nontrivial textures, such as skyrmions.[9-12] Among the few known intrinsic magnetic topological insulators (TIs), $MnBi_2Te_4$ (MBT) has been recognized for its potential to explore new correlated topological quantum states and systems.[2,13-15] MBT is the first reported topological insulator with an antiferromagnetic (AFM) ground state below its Néel temperature ($T_N \approx 25$ K). Owing to the compensated magnetic sublattice moments, MBT has no stray fields and thus promoting fast spin dynamics and insensitivity to magnetic perturbations in spin devices.[2]

Pioneering efforts have been made to study the real-space topological structures in MBT,[16] yet the local influence of topology on the creation of topological defects has not been investigated. Recently, distinguishable MBT domain walls (real-space topology) were imaged, and their dynamics has been studied, e.g. by magnetic force microscopy.[17] It has been reported that the MBT domain walls disappear around the canted antiferromagnetic (CAFM) phase boundary[17] and the AFM-CAFM phase transition area could be divided into sub-phases: antiparallel surface spin-flop states ($SSF_A$) and parallel surface spin-flop states ($SSF_P$).[16] These studies show that there may be potential magnetic structure evolution corresponding to this transition state, whereas no direct evidence has been reported. Since MBT was proposed as a possible platform for realizing skyrmions,[2,13-15] studies mainly focused on strategies such as doping with magnetic impurities[18,19] and coupling to magnetic substrates.[20,21] However, owing to the strict experimental conditions (e.g. low temperature and external magnetic field) and the imaging challenges because of compensated AFM magnetic moments, there is no easy way for the direct visualization of such topological features in antiferromagnetic MBT crystals.

Moreover, the unique surface electronic states of the TI could enable either current or electric field driven manipulation of the real-space topology in such a material, similar to recently shown control of topological protection[22], which has significant potential for topological nanoelectronics and technological applications[23,24]. Control of magnetic components in memory storage devices through precise manipulation of such functional structures has been discussed in the context of higher energy efficiency and faster information processing[25]. However, due to the complex spin and charge interactions in these systems, the underlying quantum-mechanical processes behind these phenomena are still not fully understood and further advanced theory and experiments, such as the combination of imaging and theoretical modelling techniques are required to provide further details on the competing spin, charge, and orbital interactions in these fascinating materials[26].

Here, we report the direct visualization of magnetic structure evolution at topological defects in bulk antiferromagnetic $MnBi_2Te_4$ crystals using cryogenic MFM with an external magnetic field. The domain walls exhibit an instability at the AFM-CAFM phase transition area, along with the formation of dot-shaped magnetic structures. This phenomenon is akin to the recently observed formation of merons on AFM domain walls[27], but the difference is that in our case no current pulses are needed to have this topological transition, as the transition is induced by a changing magnetic field. The key point here is that domain walls are inherently topologically nontrivial along the transverse direction, making them a natural site for further topological transformations. While skyrmions may emerge from domain walls breaking in ferromagnetic systems due to current pulses[28], our AFM system exhibits a further breakdown into 3D topological magnetic textures — chiral bobbers[29-31]. This process, dubbed the topological proximity effect, occurs because bobbers form near pre-existing topological domain walls. We find a strong phase contrast state in MFM images of these magnetic spin textures. A new magnetic phase diagram of MBT is proposed according to our MFM results, which shows regions with additional magnetic textures. We further explored the observed magnetic textures using atomistic spin simulations and visualized their three-dimensional spin structure. Our analysis suggests the formation of chiral bobbers in the bulk, which terminate at the surface with a skyrmion, following the vanishing of the domain walls in MBT crystals, via the topological proximity effect.

**Results and Discussion**

$MnBi_2Te_4$ (MBT) single crystals have a rhombohedral structure (space group $R\bar{3}m$)[32] with seven-atom-thick septuple layers (SL: Te-Bi-Te-Mn-Te-Bi-Te) as building blocks, as shown in **Fig. 1a and 1b**. Such a uniaxial A-type AFM phase is characterized by alternating ferromagnetic (FM) Mn order in adjacent septuple layers, which renders a FM intralayer exchange coupling and an AFM interlayer coupling with an out-of-plane easy axis. The magnetic phase diagram obtained by DC superconducting quantum interference device (SQUID) measurements in **Fig. S1** (Supplementary Information) confirms that our MBT single crystal has the A-type antiferromagnetic phase (AFM) below the Néel temperature ($T_N \approx 25$ K), in agreement with reported *H-T* phase diagrams[17]. The phase diagram shows temperature and magnetic field dependent phase transitions from an AFM phase to a canted-antiferromagnetic (CAM) phase and then to a paramagnetic phase (PM). The selected area electron diffraction (SAED) patterns in **Fig. 1c** show the top-view crystallinity along the c-axis of the pure MBT single crystal. A high-angle annular dark-field (HAADF) scanning transmission electron microscopy (STEM) image (**Fig. 1d**) depicts the top-view atomic structure of the MBT single crystal and includes the atomic structure model.

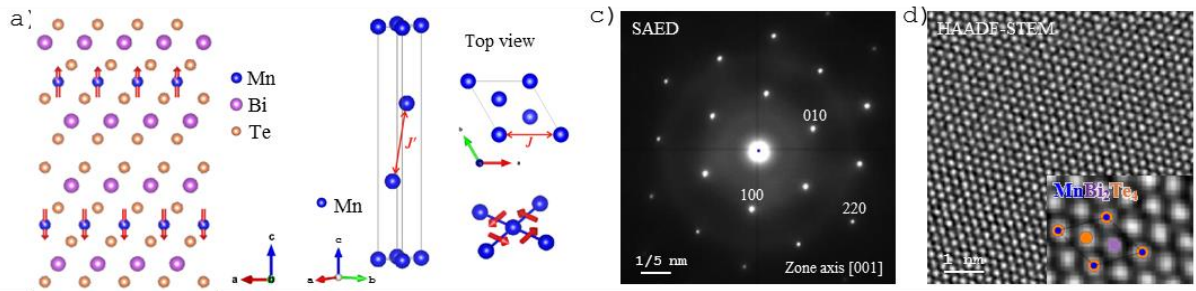

**Figure 1. MnBi$_2$Te$_4$ bulk crystal.** a) crystal structure model, b) atomic model of antiferromagnetism, c) selected area electron diffraction pattern (SAED) and d) high-angle annular dark field (HAADF) of top-view MBT bulk crystal, where the Mn, Bi and Te atoms are clearly distinguishable.

MFM measurements were performed by a dual-pass mode with a lift height of 50 nm, i.e. the MFM image was taken following the topography contour with a constant lift height of h=50 nm between the tip and the sample, as shown in the schematic in **Fig. 2a**. An applied magnetic field can modify the spin configuration at the domain wall to induce a larger out-of-plane magnetic moment and therefore a stronger DW contrast.[17] MFM detects the change of the phase signal of the cantilever oscillation, which is proportional to the stray field gradient in the out-of-plane direction. Therefore, the dark/bright contrast represents the attractive/repulsive magnetization, indicating a parallel/antiparallel magnetization of the sample in comparison to the magnetic tip moment. MBT, known as an intrinsic A-type AFM with uniaxial anisotropy, possesses two possible AFM domain states: up-down-up-down (↑↓↑↓) or down-up-down-up (↓↑↓↑).[16,17,33] These domains can be translated into each other through time reversal symmetry or fractional lattice translation, thereby being antiphase domains separated by the AFM DWs that serve as antiphase boundaries.[17] As a result, domains are either continuous or form loops in the absence of any vertex points. The clear contrast observed at the DW region arises from the susceptibility difference between the A-type AFM domain and CAFM states, known as a susceptibility contrast mechanism.[17,33] A line profile can be seen in **Fig. 2b** where a clear drop of the phase signal occurs. Such antiphase domains form a net out-of-plane moment and high magnetic susceptibility due to their spin configuration under magnetic field. The observed domain wall width is around 300-500 nm. **Fig. 2c** displays the evolution of the magnetic structure of MBT at 12 K at various magnetic fields. At 3 T the domain walls are visible as dark lines in the MFM phase image. The domain wall width increases with increasing magnetic field due to the Zeeman energy gain. When the magnetic field increases to 3.10 T, localized magnetic structures start nucleating around the magnetized DW, forming dot-like patterns with strong MFM phase contrast. With the increase of the magnetic field, the number of the dot-like structures significantly increases with a concomitant vanishing or replacement of the DWs. At a high magnetic field of 3.20 T, the magnetization is saturated resulting in diminished MFM contrast. **Fig 2d** shows the detailed magnetic feature change from antiferromagnetic domain walls under increasing magnetic field at 10 K and the transition to black dot-like features.

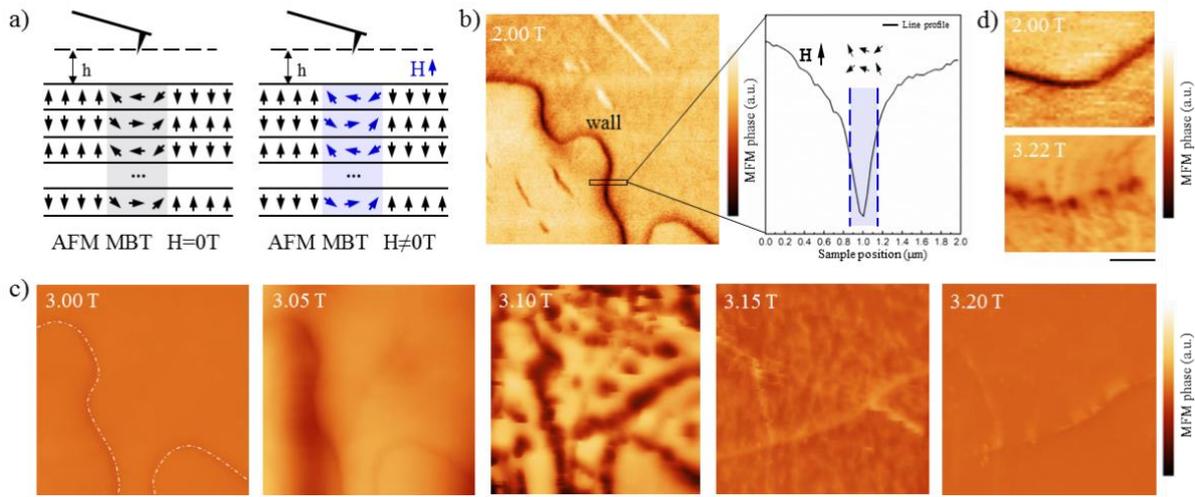

**Figure 2. Domain walls in MBT and transformation to dot-like structures with magnetic field.** a) Schemes of MFM measurement on antiferromagnet MBT domain wall without magnetic field and with magnetic field, b) MFM image and line profile of AFM MBT domain wall under 2.00 T at 12 K, c) MFM images showing the contrast change of magnetic features under increasing magnetic field at 12 K. Note that all images have the same phase contrast colour scale, i.e. magnetic contrast is especially strong at the transition point of 3.10 T, first image shows original position of domain walls by white dotted lines, d) MFM images showing the detailed magnetic feature change from antiferromagnetic domain walls under increasing magnetic field from 2.00 T to 3.22 T at 10 K and the transition to black dot-like features. Scale bar 1 µm.

**Fig. 3** shows the H-T phase diagram with data points from MFM measurements in magnetic fields from 1.0 to 4.0 T, at various temperatures from 6 K to 30 K (see **Fig. S2**). We observed the changes of magnetic features with a fine-step increase in magnetic fields around the AFM phase boundary and found two distinguishable areas at the original AFM boundary (blue and purple area). Along the AFM boundary, marked by blue dots, AFM domain walls (black curved lines in MFM images) begin to break, forming rounded dot-like features around the broken domain walls (**Fig. 2d** and MFM images in **Fig. 3a**). This indicates that the MBT sample is undergoing a surface spin-flop (SSF) transition, which may energetically favor AFM DWs as the initial transition points. We found that there is a maximum contrast value of the magnetic signal along the phase boundary, shown with purple data points around the middle of the phase transition region in **Fig. 3a**. The magnetic features with extremely large contrast are mainly the round black dot-like features with a broadened size. This may indicate that the SSF transition of AFM MBT turns from the antiparallel surface spin-flop states ($SSF_A$) into parallel surface spin-flop states ($SSF_P$)[16], possibly occurring at the center of the dot-like features. These features exhibit high contrast due to the parallel magnetization of the tip and the out-of-plane surface spin component. We also find strong interaction between the sample surface and the tip around the large contrast state, shown by the inconsistency of magnetic features along scan lines or signal jumps in MFM images. The magnetic features become broadened afterwards showing a lowered magnetic signal in MFM images (e.g., 2.80 T at 18 K and 2.70 T at 20 K in **Fig. S3**). This may indicate a metastable spin-flop state occurring on a large scale between possible $SSF_P$ states and bulk spin-flop states (BSF) in the purple region of the phase diagram. In the canted-AFM state region, no identical magnetization feature contrast was observed.

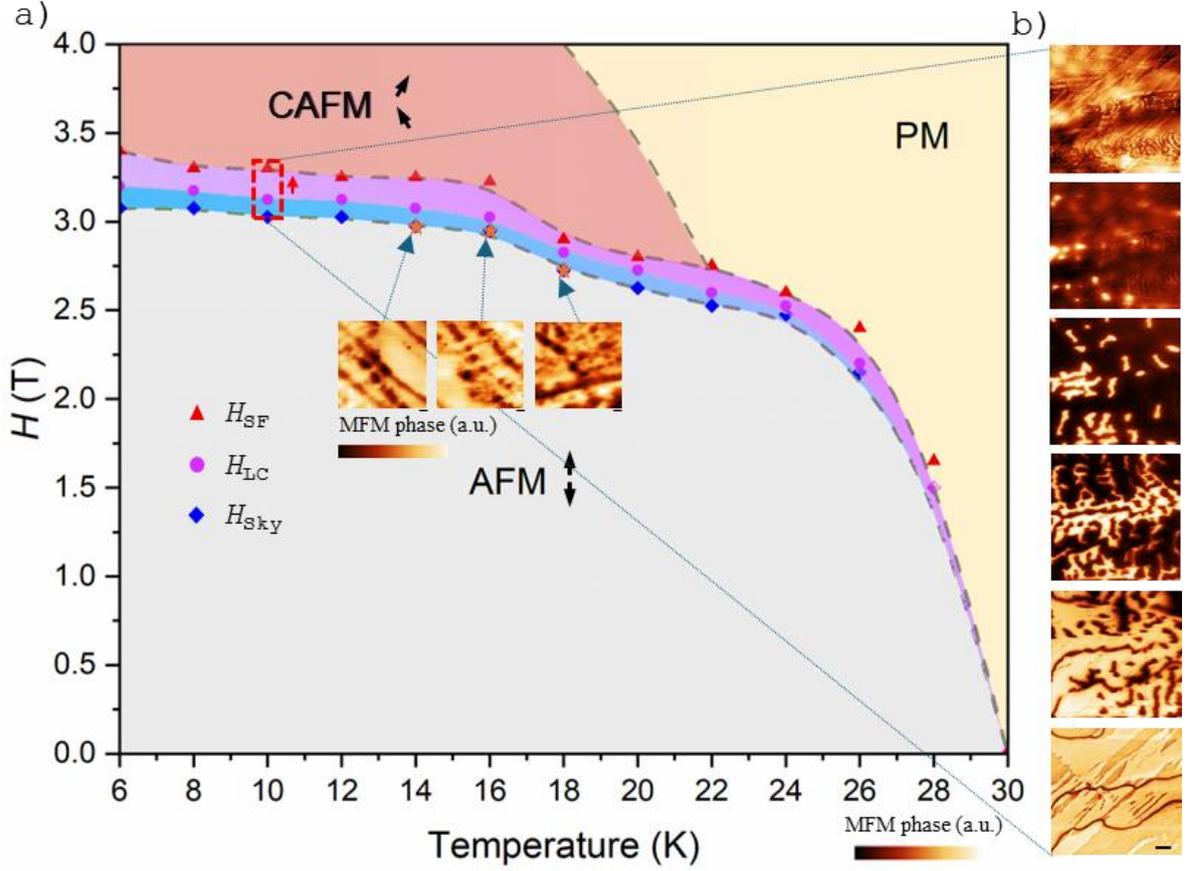

**Figure 3. MFM-derived phase diagram.** a) regions with additional magnetic textures between AFM and CAFM/PM phases (blue region with dot-like structures, purple region with strongest MFM contrast and contrast reversal in MFM; $H_{SF}$= H where spin flop, $H_{LC}$= H at large contrast area, $H_B$ = H where bobbers form), MFM images with dot-like features appearing at 14 K, 16 K and 18 K are shown in inset images, scale bar 100 nm. AFM, antiferromagnetic phase; CAFM, canted-AFM phase; PM, paramagnetic phase. b) series of MFM images of magnetic feature changes and contrast reversal within the strong contrast region with increasing magnetic field at 10 K, phase contrast colour scales vary in each image, scale bar 1 μm.

The fact that the new transition region is only seen in MFM measurements (probing the MBT crystal surface) but not SQUID measurements (probing the whole, significantly larger crystal volume) confirms the surface nature of the observed spin textures. To understand experimental observations, we employed an atomistic spin model[34]. MnBi$_2$Te$_4$ has a $R\bar{3}m$ crystal structure, characterized by the hexagonal unit cell, which is spanned by the lattice vectors, **a** = (1, 0, 0), **b** = (−1/2, √3/2, 0), **c** = (0, 0, 9.44). There are three magnetic Mn atoms in the unit cell, see **Fig. 1(b)**, whose fractional coordinates are given by Mn = (0, 0, 0), (2/3, 1/3, 1/3), (1/3, 2/3, 2/3). The spin Hamiltonian for MnBi$_2$Te$_4$ reads

$$H = -\frac{J}{2}\sum_{\langle ij \rangle \perp} \mathbf{S}_i \cdot \mathbf{S}_j - \frac{J'}{2}\sum_{\langle ij \rangle \parallel} \mathbf{S}_i \cdot \mathbf{S}_j + \sum_{\langle ij \rangle \perp} D(z)(\hat{z} \times \mathbf{r}_{ij})(\mathbf{S}_i \times \mathbf{S}_j) - K\sum_i (S_i^z)^2 - \sum_i \mu_S \mathbf{B} \cdot \mathbf{S}_{ij},$$

(1)

where $S_i$ is the normalized spin vector localized at $r_i$, $J$ ($J'$) is the in-plane (out-of-plane) nearest neighbour exchange interaction constant, $D(z)$ is the layer dependent interfacial DMI constant, $r_{ij}$ is the normalized directional vector connecting the sites $i$ and $j$, $K$ is the perpendicular (along z-axis) anisotropy constant, $\mu_S$ is the magnetic moment of each atom, and $\boldsymbol{B}$ is the magnetic field applied along z-axis. Note that MnBi$_2$Te$_4$ has inversion symmetry, which causes the bulk DMI to vanish. Therefore, the DMI should be localized at a surface. As shown in **Fig. 4(b)**, we assume that the DMI decays exponentially, $D(z) = D_0 \exp(-z/\xi)$, where $D_0$ is the DMI constant at the surface, $z$ is a distance from the surface, and $\xi$ is the localization length. The steady spin configurations are obtained by the Monte Carlo (MC) method. We employ the simulation parameters obtained from DFT calculations[35,36]; $J$ = 1.47 meV, $J'$ = 0.36 meV, $K$ = 0.23 meV, $D_0$ = 0.4 meV, $\xi$ = 2.5 nm, and $\mu_S$ = 4.04$\mu_B$, where $\mu_B$ is the Bohr magneton. To obtain the phase diagram in **Fig. S1**, we start from a uniform antiferromagnetic state at the lowest temperature (T = 0.5 K) and raise the temperature for each magnetic field. At each temperature, we thermalize the system for 10$^4$ MC steps with the periodic boundary conditions along x and y directions, while imposing an open boundary condition along z-axis.

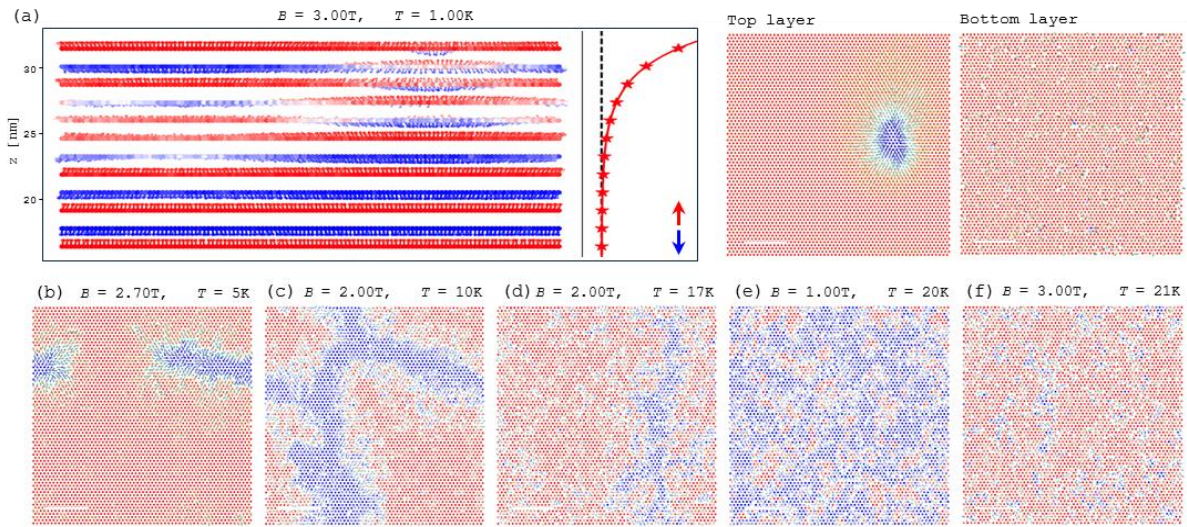

**Figure 4. Magnetic texture formation in an atomistic spin model.** a) Simulation of spin textures for the top six layers of MBT for 3.0 T and 1K: side-section view of the spin structure (left), top view of the spin structure in the top (middle) and bottom (right) layers. b-f) Top view of the simulation results for spin textures at 2.70 T and 5 K, 2.00 T and 10 K, 2.00T and 17 K, 1.00 T and 20 K, 3.00 T and 21 K, respectively (see also supplementary Fig. S4).

**Fig. 4(a)** shows that a skyrmion is stabilized in the top layer, while its size first expands and then shrinks, eventually becoming a Bloch point as moving away from the surface into the bulk, thus the entire spin texture forms an AFM chiral bobber. These bobbers appear in the vicinity of the phase boundaries, as shown in **Fig. S4(b)**. The presence of the AFM bobbers (which correspond to first expanding and then shrinking skyrmions in the top five surface layers, see **Fig. 5**) is represented by the average topological charge of these layers, as shown by the blue density plot in **Fig. S4(b)**. This behaviour is consistent with the experimental observations. The topological charge $q(z)$ is evaluated as the sum of solid angles spanned by neighbouring three spins.[37] **Fig. 4(b-f)** shows typical spin configurations at the surface layer, each corresponding position in the phase diagram is labelled in **Fig. S4(b)**. At low temperatures, an isolated skyrmion is observed, while, as temperature increases, a skyrmion grows in its size until it eventually forms a domain wall. Further increasing temperature close to the Curie temperature, the effect of thermal fluctuations becomes large and results in an unstable short-pitch structure[38], which contributes to the large average topological charge. The three-dimensional spin texture of the antiferromagnetic bobber, including an internal cross-section, is shown in **Fig. 5**.

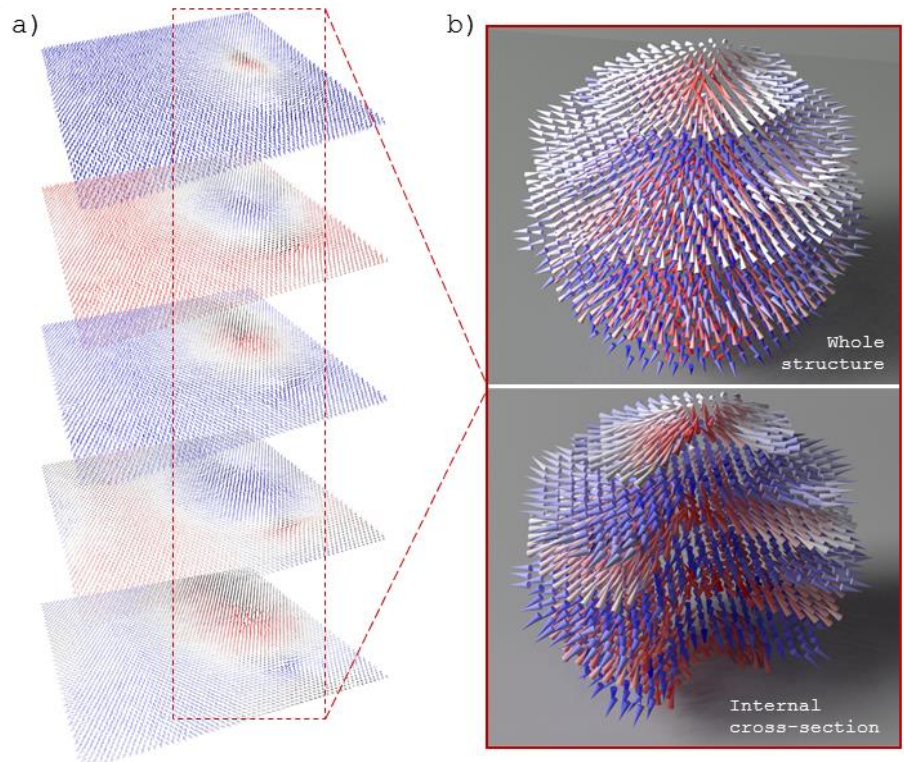

**Figure 5. Antiferromagnetic chiral bobber.** a) magnetic spin textures separately in each of top five layers, b) three-dimensional representation of spin texture in the top five layers of the MBT film shown in Fig. 4(a).

**Conclusion**

In conclusion, we have presented evidence of a real-space topological proximity effect in an antiferromagnetic topological insulator. This effect leads to the formation of nontrivial 3D topological textures at antiferromagnetic domain walls, which eventually replace the domain walls as the magnetic field increases at the AFM-CAM transition in MBT. Moreover, Monte Carlo simulations indicate the presence of AFM chiral bobbers, which are true 3D antiferromagnetic spin textures terminating with an AFM skyrmion at the MBT surface and with a Bloch point in the bulk. The existence of these chiral bobbers is highly significant for AFM spintronics applications. Their fast dynamics, robustness against external magnetic fields, and unique spin configurations offer a promising platform for energy-efficient spintronic devices. The manipulation and electric control of these AFM bobbers could enable new functionalities, such as high-speed data storage, and potentially lead to the development of antiferromagnetic memory and logic devices — advancements that are essential for the next generation of spintronic technologies. Our findings point to the opportunity of exploiting functional real-space topology and topological proximity effects in a topological insulator system.[29]

**Experimental section**

*SQUID measurement:* The magnetic properties of $MnBi_2Te_4$ single crystals were measured using a Superconducting Quantum Interference Device (SQUID) magnetometer (Quantum Design MPMS). Isothermal magnetization curves (M-H loops) were acquired in the temperature range of 5-300 K under applied magnetic fields ranging up to 8 T to evaluate the hysteresis behavior. All measurements were performed under high-vacuum conditions ($<10^{-5}$ mbar) to suppress thermal drift and minimize environmental interference.

*Magnetic Force Microscopy Characterisation:* MFM measurements were carried out in an attocube attoAFM I atomic force microscope in cryogenic magnetic force microscopy (MFM) mode equipped with a superconducting magnet (attoDRY1000). The experiments were performed with commercially available hard magnetic coated Si tips of resistivity 0.01-0.02 Ωcm (NanoSensors PPP-MFMR). MFM images were obtained with a tip lift height ~50nm at various magnetic fields from 1.0 T to 4.0 T at temperatures from 6K to 28K.

*TEM:* TEM samples were prepared by mechanical exfoliation and transfer to Cu mesh TEM grids. HAADF STEM imaging was performed with a JEOL ARM 200CF equipped with a cold field-emission source and two aberration correctors at 200 kV energy. The range of collection angles was 67-275 mrad for HAADF.

**Acknowledgements**

J.S. and O.A.T. acknowledge funding by the Australian Research Council through Discovery Grants. O.A.T. acknowledges support by the NCMAS grant and ICC-IMR visitor program from Tohoku University, Japan. The work at Brookhaven National Laboratory is supported by the U.S. DOE Basic Energy Sciences, Materials Sciences and Engineering Division under Contract No. DESC0012704. L. L. acknowledges support from the National Natural Science Foundation of China (Grant No. 12204096), the Natural Science Foundation of Jiangsu Province (Grant No. BK20220797).

# Supplementary Information

# Antiferromagnetic Chiral Bobber Formation and Topological Proximity Effect in MnBi$_2$Te$_4$

Y. Xu[1], D. Kurebayashi[2], D. Zhang[1], P. Schoenherr[1], L. Li [3], Z. Yue[4] *, W. J. Ren [5], M.-G. Han[6], Y. Zhu[6], Z. Cheng[7], X. Wang[7], Oleg A. Tretiakov[2] *, and J. Seidel[1] *

[1] *School of Materials Science and Engineering, UNSW Sydney, Sydney 2052, Australia*
[2] *School of Physics, UNSW Sydney, Sydney 2052, Australia*
[3] *Key Laboratory of Quantum Materials and Devices of Ministry of Education, School of Physics, Southeast University, Nanjing 211189, China*
[4] *School of Artificial Intelligence Science and Technology, University of Shanghai for Science and Technology, Shanghai 200093, China; Institute of Photonic Chips, University of Shanghai for Science and Technology, Shanghai 200093, China*
[5] *Shenyang National Laboratory for Materials Science, Institute of Metal Research, Chinese Academy of Sciences, Shenyang, 110016, China*
[6] *Condensed Matter Physics and Materials Sciences Department, Brookhaven National Laboratory, Upton, NY 11973, USA*
[7] *Institute for Superconducting and Electronic Materials, Institute for Innovative Materials, University of Wollongong, Wollongong, NSW 2522, Australia*
* Corresponding authors: zengjiyue@usst.edu.cn, o.tretiakov@unsw.edu.au, jan.seidel@unsw.edu.au

**SQUID magnetisation measurement**

The MBT single crystal is confirmed to have an A-type antiferromagnetic phase at zero magnetic field below 25 K by DC superconducting quantum interference device (SQUID) measurements and the resulting phase diagram:

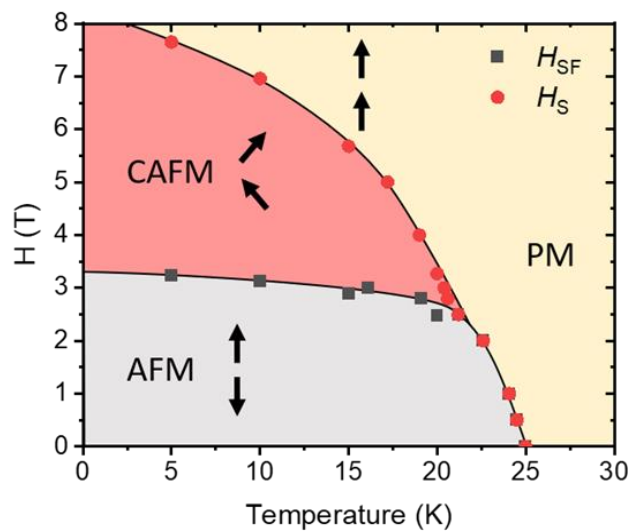

**Figure S1.** SQUID phase diagram showing the antiferromagnetic ground state below 25 K.

**MFM images at different points in the phase diagram**

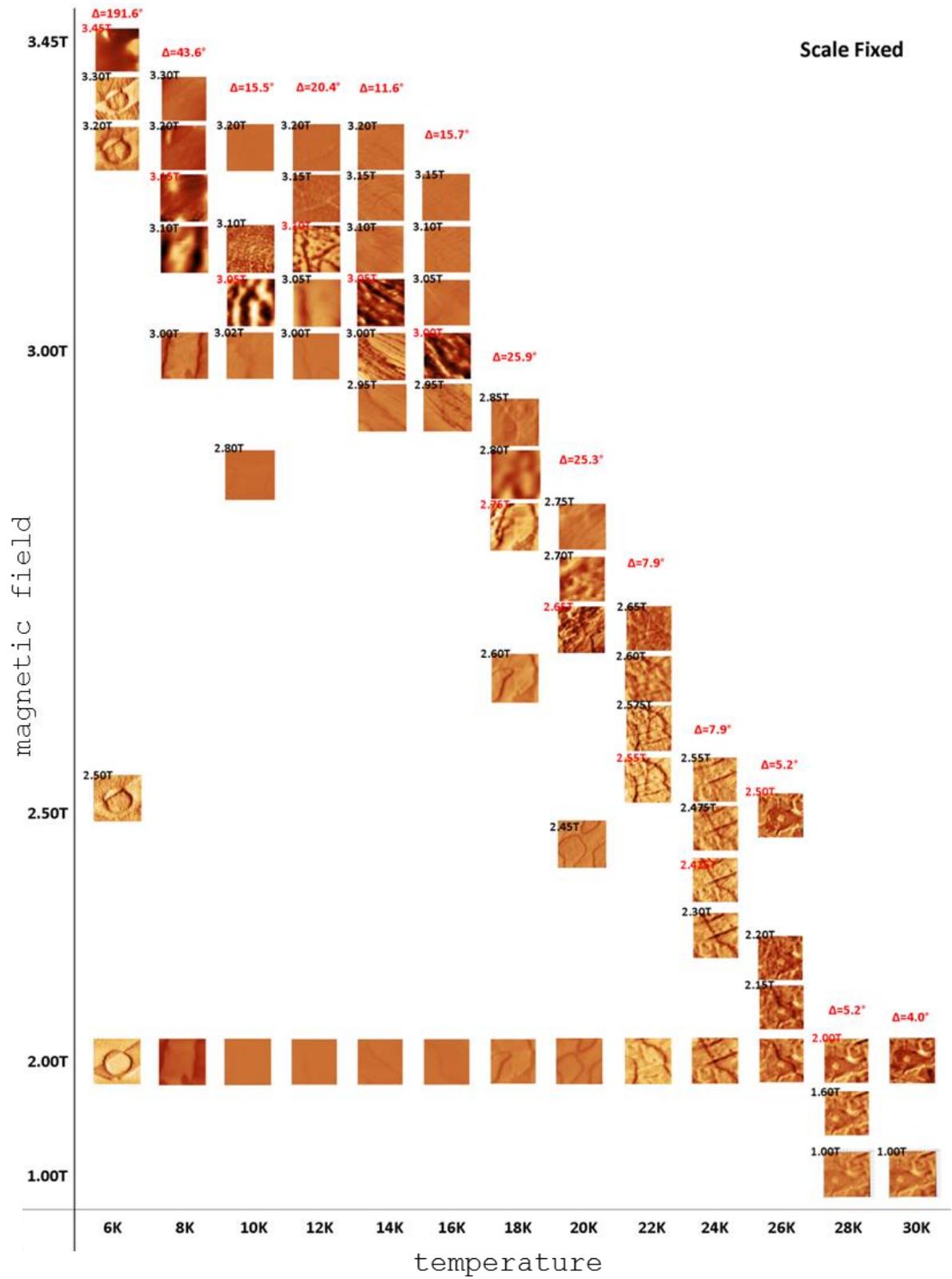

**Figure S2.** Magnetic texture evolution seen in MFM images (all images have the same MFM phase range).

## Magnetic texture evolution

The following presents MFM images at varying magnetic fields and temperatures.

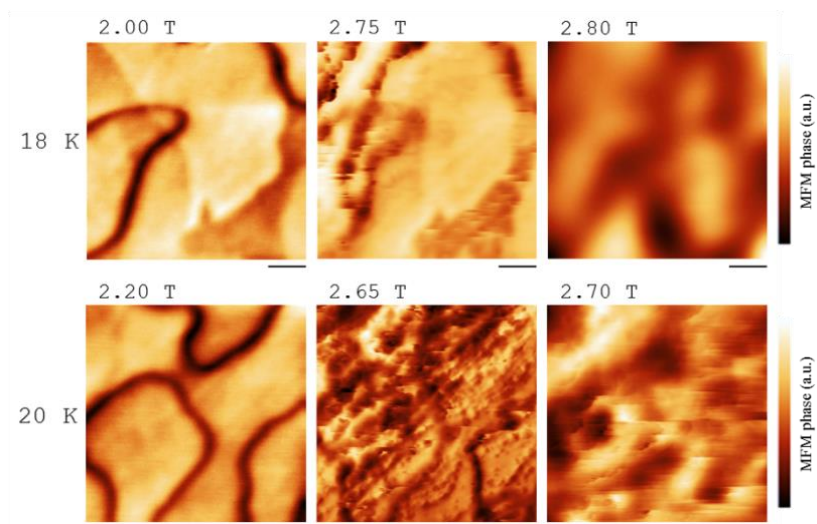

**Figure S3.** MFM images showing broadening of magnetic textures from 2.00 T to 2.80 T at 18 K and 2.20 T to 2.70 T at 20 K.

## Spin texture simulations and associated phase diagram

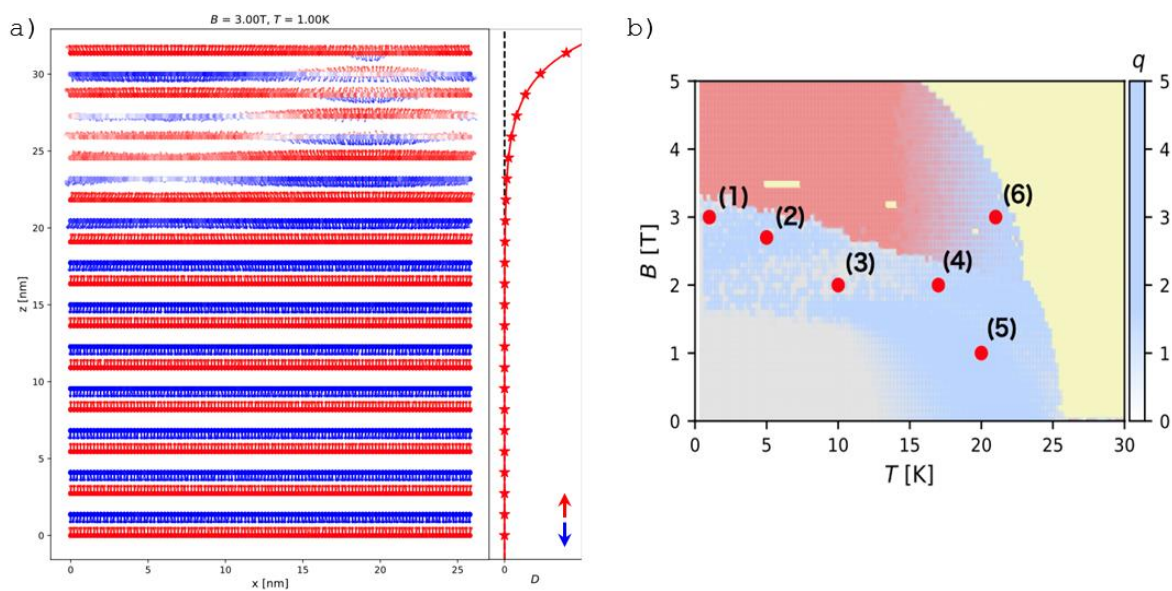

**Figure S4.** a) side-view spin layers from surface (top) into bulk (bottom) of MBT with a skyrmion of different radius in each top layer comprising the three-dimensional chiral bobber texture, b) simulated phase diagram of spin configuration, including calculated spin textures for (1) 3.0 T and 1K, (2) 2.70 T and 5 K, (3) 2.00 T and 10 K, (4) 2.00T and 17 K, (5) 1.00 T and 20 K, (6) 3.00 T and 21 K, respectively (see Fig. 4 of main text).